\documentclass[superscriptaddress, aps, prb, twocolumn, notitlepage, 10pt]{revtex4-1}
 
\usepackage{booktabs}
\usepackage{graphicx}

\usepackage{natbib}
\usepackage{amsmath}%
\usepackage{amsfonts}%
\usepackage{amssymb}%
\usepackage{array}
\pdfoutput=1
\usepackage{checkend}	
\usepackage{xspace}
\usepackage{color}
\definecolor{greytext}{gray}{0.5}

\usepackage[unicode=true,
  linktocpage,
  linkbordercolor={0.5 0.5 1},
  citebordercolor={0.5 1 0.5},
  linkcolor=blue]{hyperref}

\newcommand{\YBCO}[1]{YBa\ensuremath{_2}Cu\ensuremath{_3}O\ensuremath{_{6{.#1}}}\xspace}

\newcommand{\YBCOy}{YBa\ensuremath{_2}Cu\ensuremath{_3}O\ensuremath{_{y}}\xspace}

\newcommand{\caxis}{\ensuremath{\hat{c}}-axis\xspace}

\newcommand{\abplane}{\ensuremath{\hat{a}}-\ensuremath{\hat{b}}-plane\xspace}

\newcommand{\Tc}{\ensuremath{T_c}\xspace}
\newcommand{\highTc}{\ensuremath{\mathrm{high}\!-\!T_c}\xspace}

\newcommand{\HcOne}{\ensuremath{H_{c1}}\xspace}
\newcommand{\Hc}{\ensuremath{H_{c}}\xspace}
\newcommand{\HcTwo}{\ensuremath{H_{c2}}\xspace}

\newcommand{\Bm}{\ensuremath{B_{m}}\xspace}

\newcommand{\muSR}{\ensuremath{\mu \mathrm{SR}}\xspace}
\newcommand{\muZero}{\ensuremath{\mu_0}\xspace}

\begin{document}

\title{Vortex Lattice Melting and \texorpdfstring{\HcTwo}~in underdoped \texorpdfstring{YBa$_2$Cu$_3$O$_{y}$ }.}

\let\clearpage\relax
\author{B.~J.~Ramshaw}
\affiliation{Department of Physics and Astronomy, University of British Columbia, Vancouver, BC, Canada, V6T~1Z1}
\author{James Day}
\affiliation{Department of Physics and Astronomy, University of British Columbia, Vancouver, BC, Canada, V6T~1Z1}
\author{Baptiste Vignolle}
\affiliation{ Laboratoire National des Champs Magn\'etiques Intenses, UPR 3228, CNRS-INSA-UJF-UPS, Toulouse, France}
\author{David LeBoeuf}
\affiliation{ Laboratoire National des Champs Magn\'etiques Intenses, UPR 3228, CNRS-INSA-UJF-UPS, Toulouse, France}
\author{P.~Dosanjh}
\affiliation{Department of Physics and Astronomy, University of British Columbia, Vancouver, BC, Canada, V6T~1Z1}
\author{Cyril Proust}
\affiliation{ Laboratoire National des Champs Magn\'etiques Intenses, UPR 3228, CNRS-INSA-UJF-UPS, Toulouse, France}
\affiliation{Canadian Institute for Advanced Research, Toronto, Canada}
\author{Louis Taillefer}
\affiliation{D\'epartement de Physique and RQMP, Universit\'e de Sherbrooke, Sherbrooke, Qu\'ebec, Canada J1K 2R1}
\affiliation{Canadian Institute for Advanced Research, Toronto, Canada}
\author{Ruixing Liang}
\affiliation{Department of Physics and Astronomy, University of British Columbia, Vancouver, BC, Canada, V6T~1Z1}
\affiliation{Canadian Institute for Advanced Research, Toronto, Canada}
\author{W.~N.~Hardy}
\affiliation{Department of Physics and Astronomy, University of British Columbia, Vancouver, BC, Canada, V6T~1Z1}
\affiliation{Canadian Institute for Advanced Research, Toronto, Canada}
\author{D.~A.~Bonn}
\affiliation{Department of Physics and Astronomy, University of British Columbia, Vancouver, BC, Canada, V6T~1Z1}
\affiliation{Canadian Institute for Advanced Research, Toronto, Canada}

\date{June 29, 2012}

\begin{abstract}
Vortices in a type-II superconductor form a lattice structure that melts when the thermal displacement of the vortices is an appreciable fraction of the distance between vortices. In an anisotropic \highTc superconductor, such as \YBCOy, the magnetic field value where this melting occurs can be much lower than the mean-field critical field \HcTwo. We examine this melting transition in \YBCOy with oxygen content $y$ from 6.45 to 6.92, and we perform the first quantitative analysis of this transition in the cuprates by fitting the data to a theory of vortex-lattice melting. The quality of the fits indicates that the transition to a resistive state is indeed the vortex lattice melting transition, with the shape of the melting curves being consistent with the known change in penetration depth anisotropy from underdoped to optimally doped \YBCOy. We establish these fits as a valid technique for finding $\HcTwo(T\!=\!0)$ from higher temperature data when the experimentally accessible fields are not sufficient to melt the lattice at zero temperature (near optimal doping).  From the fits we extract $\HcTwo(T\!=\!0)$ as a function of hole doping. The unusual doping dependence of $\HcTwo(T\!=\!0)$ points to some form of electronic order competing with superconductivity around 0.12 hole doping.
\end{abstract}

\maketitle

\section{Introduction}
Cuprate high-\Tc superconductors are of great interest not only because of their high transition temperatures, but also because strong-correlation physics gives rise to peculiar normal-state properties.  Ironically, however, the strength of the superconductivity in these \highTc materials is what interferes with measurement of the normal state properties at low temperature. Applying high magnetic fields can overcome this and has led to the discovery of a small Fermi surface in underdoped \YBCOy via quantum oscillation measurements in pulsed fields. \cite{Doiron-Leyraud:2007} This discovery prompted a large experimental survey of the transport and thermodynamic properties of \YBCOy in high fields. The questions remain as to whether the high fields are revealing the normal-state properties of \YBCOy, or are instead exposing a qualitatively different field-induced ground state, or whether one might still be in a regime dominated by superconducting pairing and the presence of vortices.

The idea of high magnetic fields revealing the normal state in cuprate superconductors is a contentious one, in part because the phase diagram of the cuprates differs qualitatively from that of conventional type-II superconductors. Owing to the short coherence length, low superfluid phase stiffness, and strong anisotropy, fluctuations play a dominant role in the phase diagram. There is evidence for 3D-XY critical fluctuations below and above \Tc. \cite{Kamal:1994, Pasler:1998, Xu:2009} Previous transport measurements on several cuprate compounds have shown that reaching the resistive state requires very high magnetic fields, and that the onset of resistivity as a function of field and temperature does not follow the conventional \HcTwo curve derived from Ginzburg-Landau theory, as it does in more conventional type-II superconductors. \cite{Liang:1996, Ando:1999} Instead, as is expected for a superconductor governed by strong thermal fluctuations, a vortex melting transition occurs, \cite{Fisher:1991, Liang:1996, Safar:1992} with an extensive crossover regime to the normal state. Some Nernst effect experiments have been taken as evidence for the presence of superconducting pairing far above \Tc, even in strong magnetic fields. \cite{Wang:2006} With this in mind, it is important to consider at which field scale is superconductivity completely suppressed and the normal state recovered, especially with regard to quantum oscillation experiments which are purported to probe the ``normal state'' Fermi surface. In this paper we present, for the first time in the cuprates, a detailed comparison of the melting transition in \YBCOy with  the theory of vortex-lattice melting.

\section{Theory}
The thermodynamic critical field \Hc is the field at which superconductivity is destroyed in a type-I superconductor, and is directly related to the condensation energy of the superconducting ground state.  In a type-II superconductor the magnetic field can penetrate the sample at a field lower than \Hc. At this field, \HcOne, the magnetic field penetrates the superconductor in the form of vortices, with each vortex being supercurrent running around a normal state core and containing a quantum of magnetic flux. The cores of these vortices, whose size is of order the superconducting coherence length $\xi_0$, are in the normal state; outside of the vortex cores, the strength of the magnetic field decays over the length scale of the penetration depth $\lambda$ which, for strongly type-II superconductors such as the cuprates, is much larger than the coherence length. These vortices can form a two-dimensional lattice perpendicular to the applied field (a ``vortex lattice''), \cite{Abrikosov:1957} and the lattice spacing shrinks in size as the magnetic field is increased. As long as the vortices remain pinned, the zero-resistance property is maintained in the material. When the vortex cores overlap at a second field scale \HcTwo , superconductivity is destroyed. In an isotropic, low-\Tc type-II superconductor, such as Nb$_3$Sn, resistivity sets in at \HcTwo and the diamagnetic signal of superconductivity completely disappears. In terms of the mean-field Ginzburg-Landau coherence length $\xi_0$, this field scale is
\begin{equation}
\muZero\HcTwo(T\!=\!0) = \frac{\Phi_0}{2 \pi \xi_0^2},
\label{eq:HcTwo}
\end{equation}
where $\Phi_0$ is the flux-quantum in SI units ($\HcTwo(T\!=\!0)$ will henceforth be $\HcTwo(0)$) \footnote{All formulae in this paper have been converted to SI units. Factors of \muZero that may seem redundant, such as in \autoref{eq:Gi0} and \autoref{eq:Gi1}, have been left un-cancelled in order to keep the conversion to SI units transparent.}.

The situation is more complicated in high-\Tc materials, where the vortex lattice can melt into a vortex liquid well below \HcTwo. The Lindemann criterion for melting requires the thermal displacement of a lattice to be some fraction (defined $c_L$) of the average lattice constant.  Using the Lindemann criterion for a vortex lattice, \citet{Houghton:1989} have shown that, because of the large anisotropy in the cuprates, the vortex lattice in a strongly type-II superconductor with a high \Tc can melt at field values \Bm well below \HcTwo for intermediate temperatures (away from 0 K and \Tc). \cite{Blatter:1994} In these materials, \HcTwo represents a crossover from a vortex-liquid to the normal state.  In the traditional picture the melting field \Bm and $\muZero \HcTwo$ are equal at zero temperature, since there are no thermal fluctuations at zero temperature to melt the vortex lattice. The presence of strong quantum fluctuations could result in a vortex liquid persisting down to zero temperature. However, in order to compare our experimental data with the theory of vortex lattice melting, we use the assumption made by \citet{Houghton:1989}, \citet{Blatter:1994}, and others that $\Bm(0) = \muZero \HcTwo(0)$.

 Using the notation of \citet{Blatter:1994}, the melting transition field \Bm is given implicitly by
\begin{equation}
\frac{\sqrt{b_m(t)}}{1-b_m(t)}\frac{t}{\sqrt{1-t}}\left[ \frac{4 \left(\sqrt{2} - 1 \right)}{\sqrt{1-b_m(t)}} + 1 \right] = \frac{2 \pi c_L^2}{\sqrt{Gi}}.
\label{eq:bm1}
\end{equation}
The reduced field variable is $b_m = \Bm/ \muZero\HcTwo$, and $t =T/\Tc$ is the reduced temperature.

The Ginzburg number $Gi$, on the right hand side of \autoref{eq:bm1},  is given by
\begin{align}
Gi &= \frac{1}{2} \left( \frac{k_B T_c \gamma}{\frac{4 \pi}{\mu_0} \left(\muZero H_c(T\!=\!0)\right)^2 \xi^3_0}\right)^2 \label{eq:Gi0} \\ 
     &\approx \left(9.225 \times 10^{8} \left[\mathrm{Wb}^{-1}\mathrm{K}^{-1}\right] \times \muZero \HcTwo(0) \Tc \lambda_{ab} \lambda_c\right)^2,
\label{eq:Gi1}
\end{align}
where $\gamma$ is the anisotropy ratio $\gamma \equiv \frac{\lambda_{c}}{\lambda_{ab}}$, and the definition $\HcTwo(0) \equiv \frac{\frac{4 \pi}{\muZero} \lambda_{ab}^2 \left(\muZero H_c(T\!=\!0)\right)^2}{\Phi_0}$ has been used ($\lambda_{ab}$ and $\lambda_{c}$ are the penetration depths parallel and perpendicular to the \abplane at zero temperature). As emphasized by \citet{Blatter:1994}, this Ginzburg number should be thought of as a useful collection of parameters, and not as a number describing the width of fluctuations around \Tc as it is in more three-dimensional superconductors. The Lindemann number $c_L$ appearing on the right hand side of \autoref{eq:bm1} represents the fraction of the vortex lattice parameter, $a_v \equiv \sqrt{\frac{\Phi_0}{B}}$, that the thermal displacement must reach in order for the vortex lattice to melt. \cite{Houghton:1989, Fisher:1991,Blatter:1994}  Attempts have been made to calculate $c_L$, with values between 0.2 and 0.4 obtained for the cuprates, depending on the specific model (see \citet{Blatter:1994} for a review), but $c_L$ is probably better left as a fit parameter. \cite{Kim:1996}

\section{Experiment}

\begin{figure}%
\includegraphics[width=\columnwidth, natwidth = 38.81cm, natheight = 24.09cm]{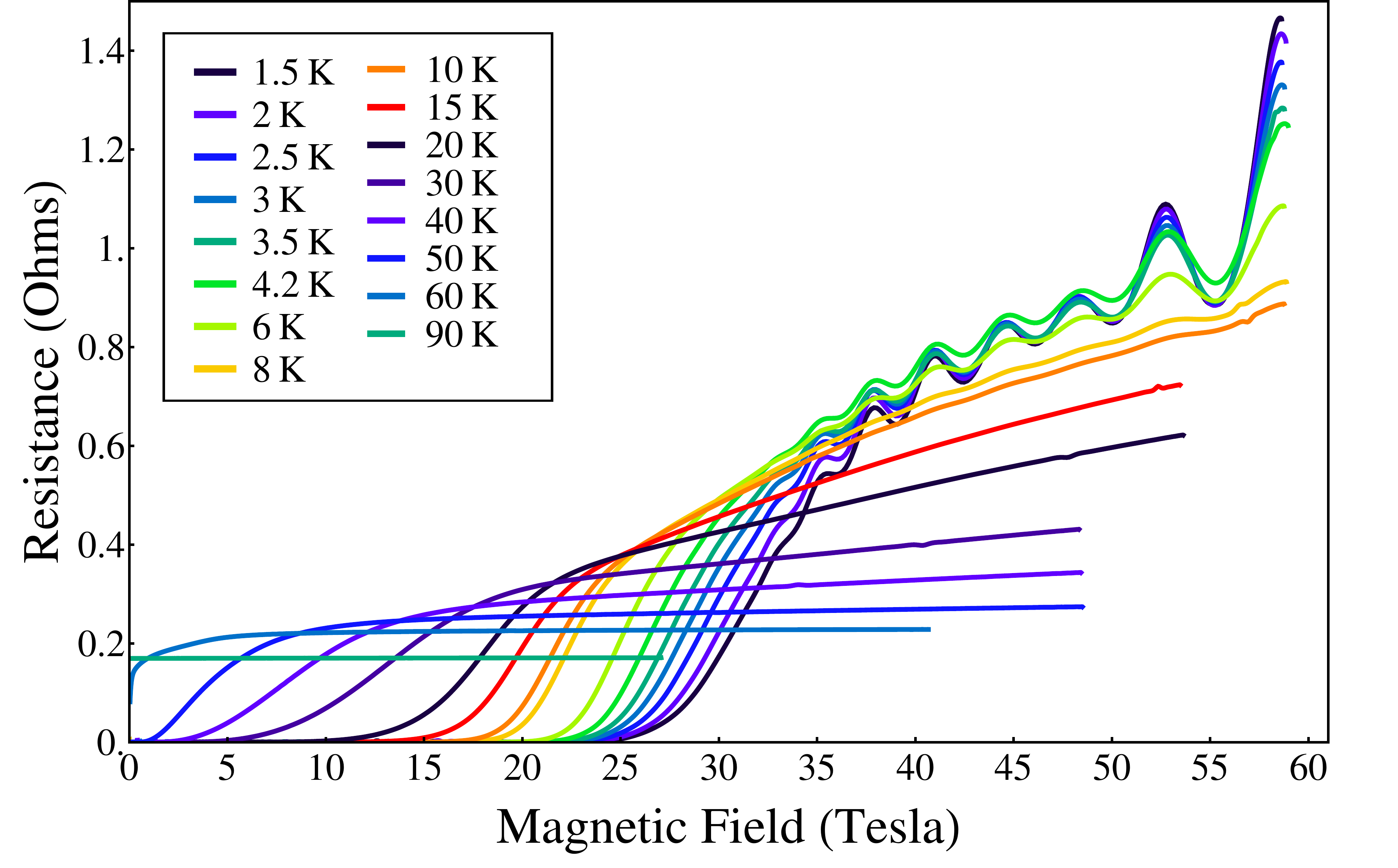}
\caption{The \caxis resistance of \YBCO{59} as a function of magnetic field, from 1.5 to 200 K. The onset of resistivity as field is increased marks the vortex lattice melting transition. At low temperatures, quantum oscillations are seen above this melting field.}
\label{fig:6p59rvb}
\end{figure}

\begin{figure}%
\includegraphics[width=\columnwidth,  natwidth = 38.81cm, natheight = 23.78cm]{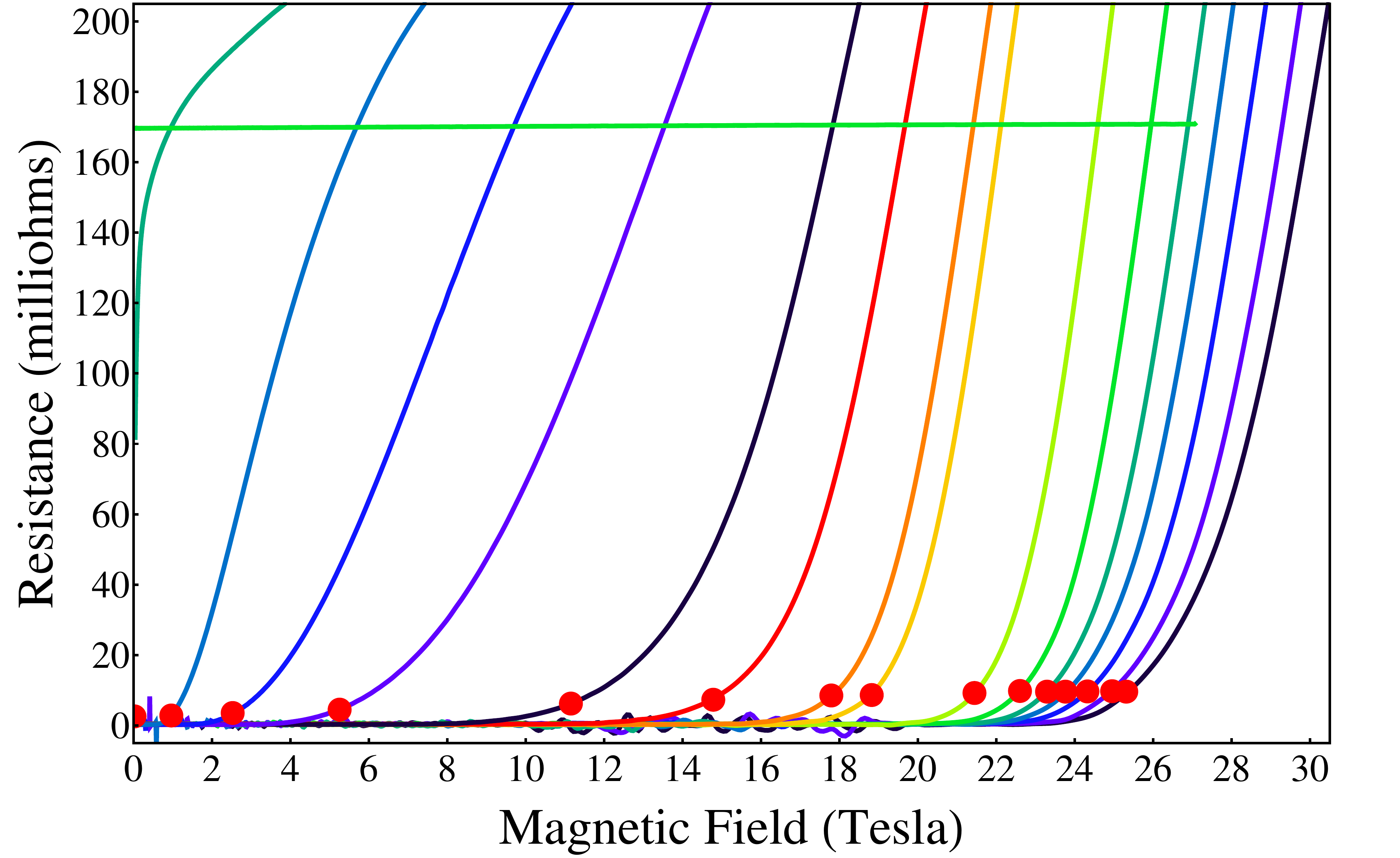} \\
\includegraphics[width=\columnwidth,  natwidth = 38.81cm, natheight = 23.78cm]{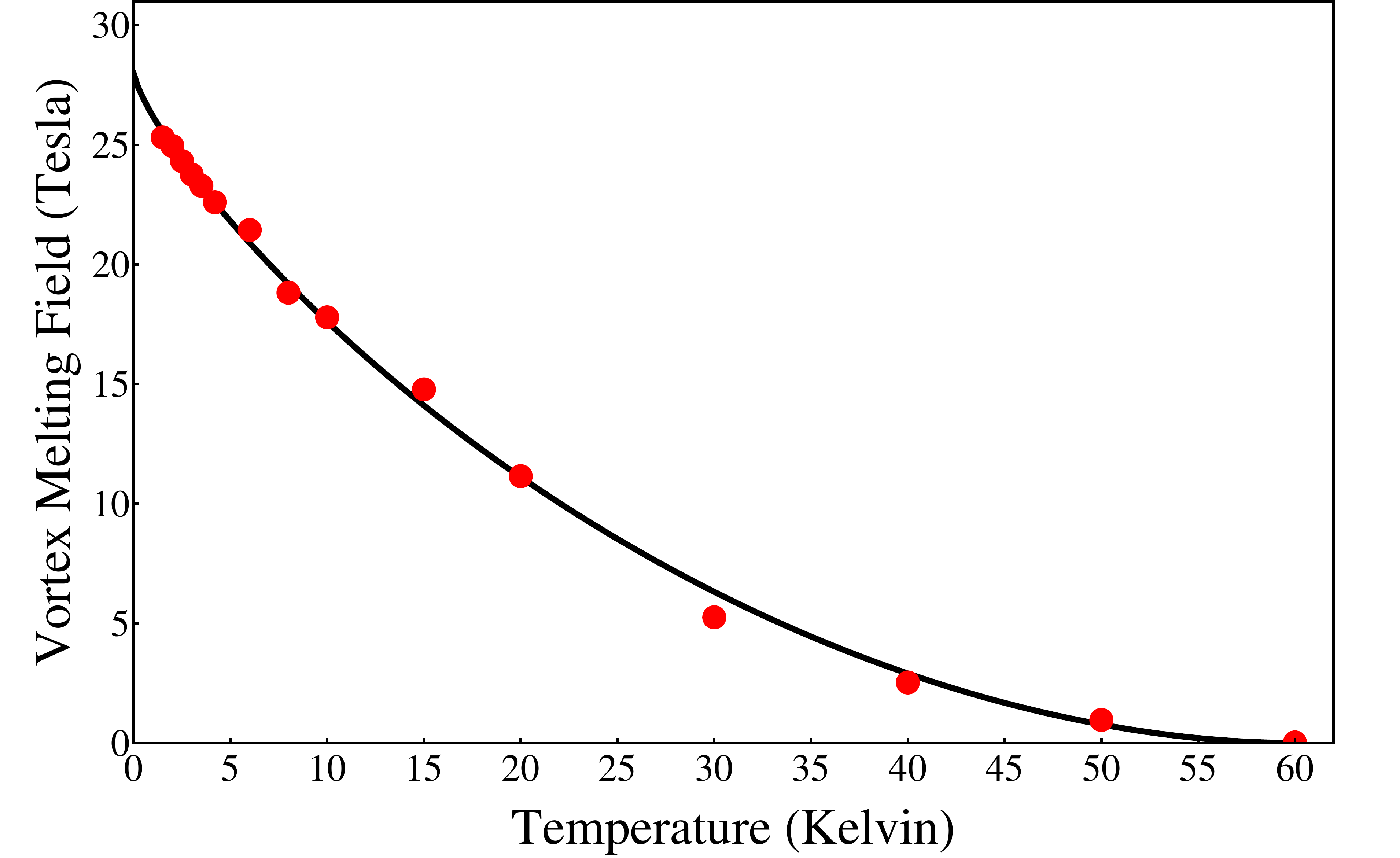}
\caption{\textit{Top panel:} A magnified plot of the resistive transitions shown above in \autoref{fig:6p59rvb}. The red dots are where the resistance is $1/100^{th}$ of its value at 60 Telsa (extrapolated for high temperatures where resistance was not measured to the highest fields).  \textit{Bottom panel:} The same data points as highlighted in red in the top panel, now plotted as a function of temperature. The black line is a fit is to \autoref{eq:bm1}, using the known parameters given in \autoref{tab:correlation lengths}, and gives $\muZero \HcTwo(0) = 28 \pm 0.3$~T, and $c_L = 0.37$.}%
\label{fig:6p592plots}%
\end{figure}

All of the samples used in this study were fully-detwinned, single-crystal \YBCOy, grown in barium zirconate crucibles and annealed in oxygen to the desired concentration.\cite{Liang:1998} Gold contacts were evaporated onto the a-b faces for a four-point c-axis resistivity geometry, and the gold was partially diffused into the sample near $500^{\circ}$C to obtain sub-ohm contacts.\cite{Ramshaw:2012} The chain oxygen was then ordered into superstructures (ortho-II for \YBCO{45} through \YBCO{59}, ortho-VIII for \YBCO{67}, ortho-III for \YBCO{75}, and ortho-I for \YBCO{86} and \YBCO{92}) by annealing the samples just below the superstructure transition temperature.\cite{Zimmermann:2003}  \autoref{fig:6p59rvb} shows a typical set of \caxis resistivity curves up to 60 tesla, from 1.5 to 200 K for \YBCO{59}. We define the resistive vortex-melting transition as the magnetic field where the resistance is $1/100^{th}$ of its value at 60 tesla. The definition of \Bm from resistivity curves is somewhat uncertain because of the width of the resistive transition (see upper panel of \autoref{fig:6p592plots}). An alternative definition would be the intersection of a line tangent to the steepest part of the resistive transition with the temperature axis. This would lead to small offsets (one tesla at most) in \Bm, but would not otherwise affect the conclusions of this paper. However, it is important that a consistent definition across different doping levels be used. 

\begin{figure}%
\includegraphics[width=\columnwidth,natwidth = 38.81cm, natheight = 25.24cm]{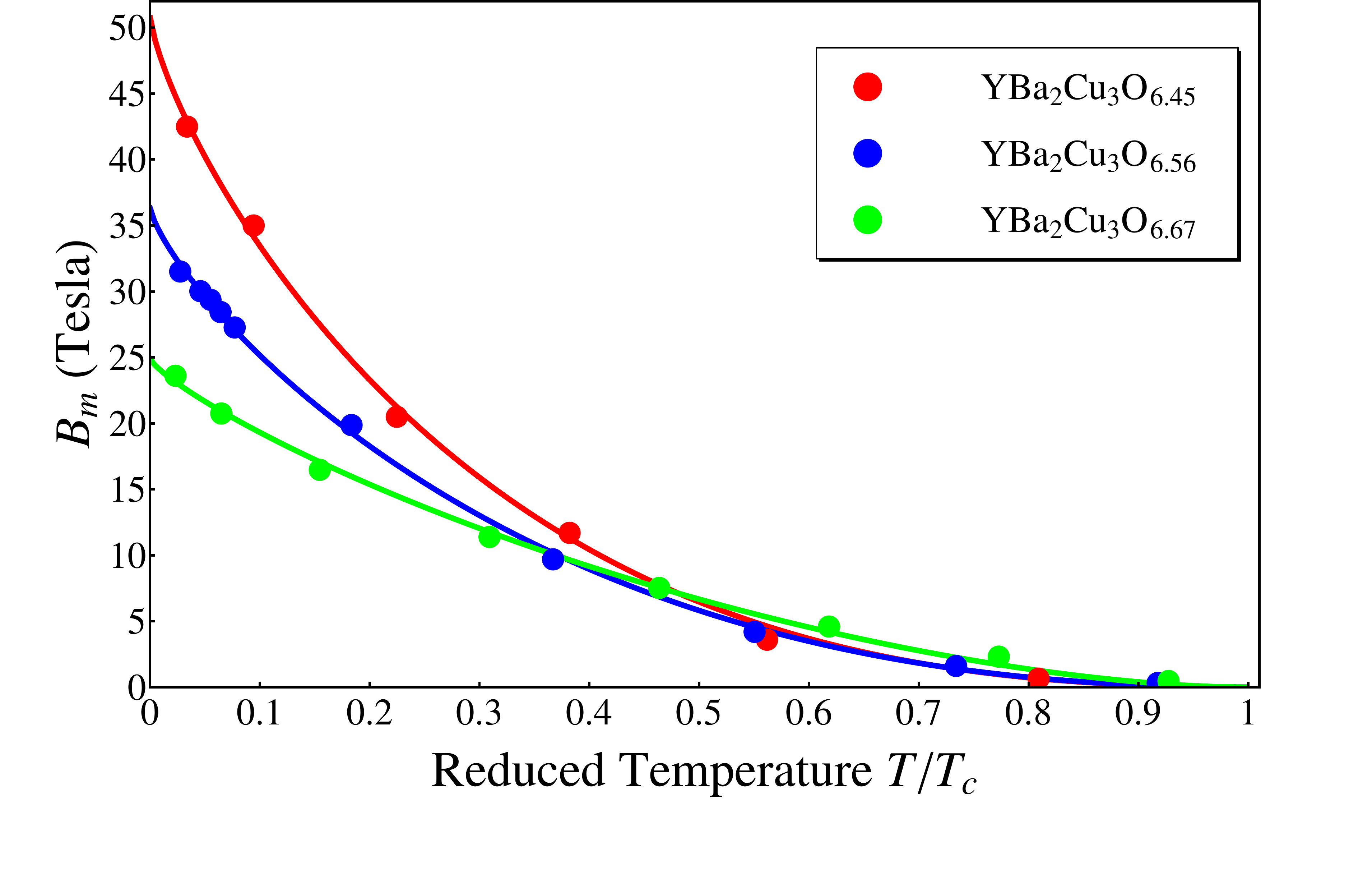} \\
\includegraphics[width=\columnwidth,natwidth = 38.81cm, natheight = 25.27cm]{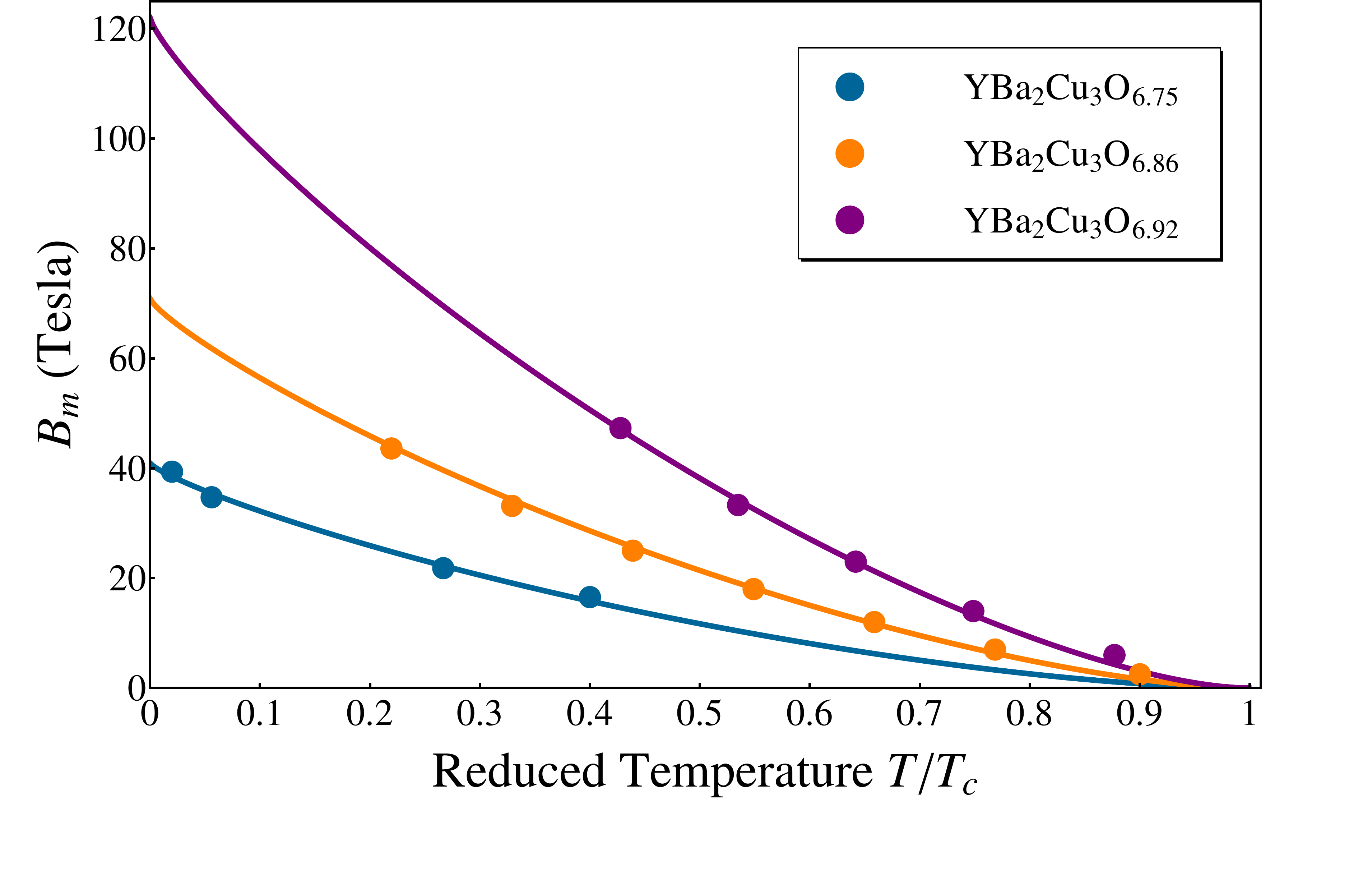}
\caption{The vortex lattice melting transition for several different oxygen concentrations. The temperature axis has been scaled by \Tc, and the lines are best-fit lines to \autoref{eq:bm1}. All of the data points were aquired in the same manner as described in the caption of \autoref{fig:6p592plots}.}%
\label{fig:bvts}%
\end{figure}

The upper panel of \autoref{fig:6p592plots} shows the vortex lattice melting transition from 1.2 K up to \Tc for \YBCO{59}, one of the underdoped samples in which the melting transition is accessible even at low temperatures. The concave upwards shape is characteristic of a vortex melting transition, as seen before in \YBCOy and in other cuprates, \cite{Ando:1999, Kim:1996} and differs qualitatively from the concave downwards curvature of $H_{c2}(T)$ in conventional superconductors. This form has been observed in a number of cuprates, \cite{Ando:1999, Lake:2002, LeBoeuf:2011} but a systematic comparison to \autoref{eq:bm1} across the underdoped regime of the cuprates has not been performed. Here we present data for \YBCOy from oxygen content 6.45 to 6.92, with $T_c$s ranging from 44.5 to 93.5 K, and identify trends that arise as a function of doping. Characteristic curves for several other dopings are shown in \autoref{fig:bvts}, all with an upwards curvature, although that shape becomes less pronounced for the higher \Tc samples.

\begin{figure}%
\includegraphics[width=\columnwidth,natwidth = 38.81cm, natheight = 23.62cm]{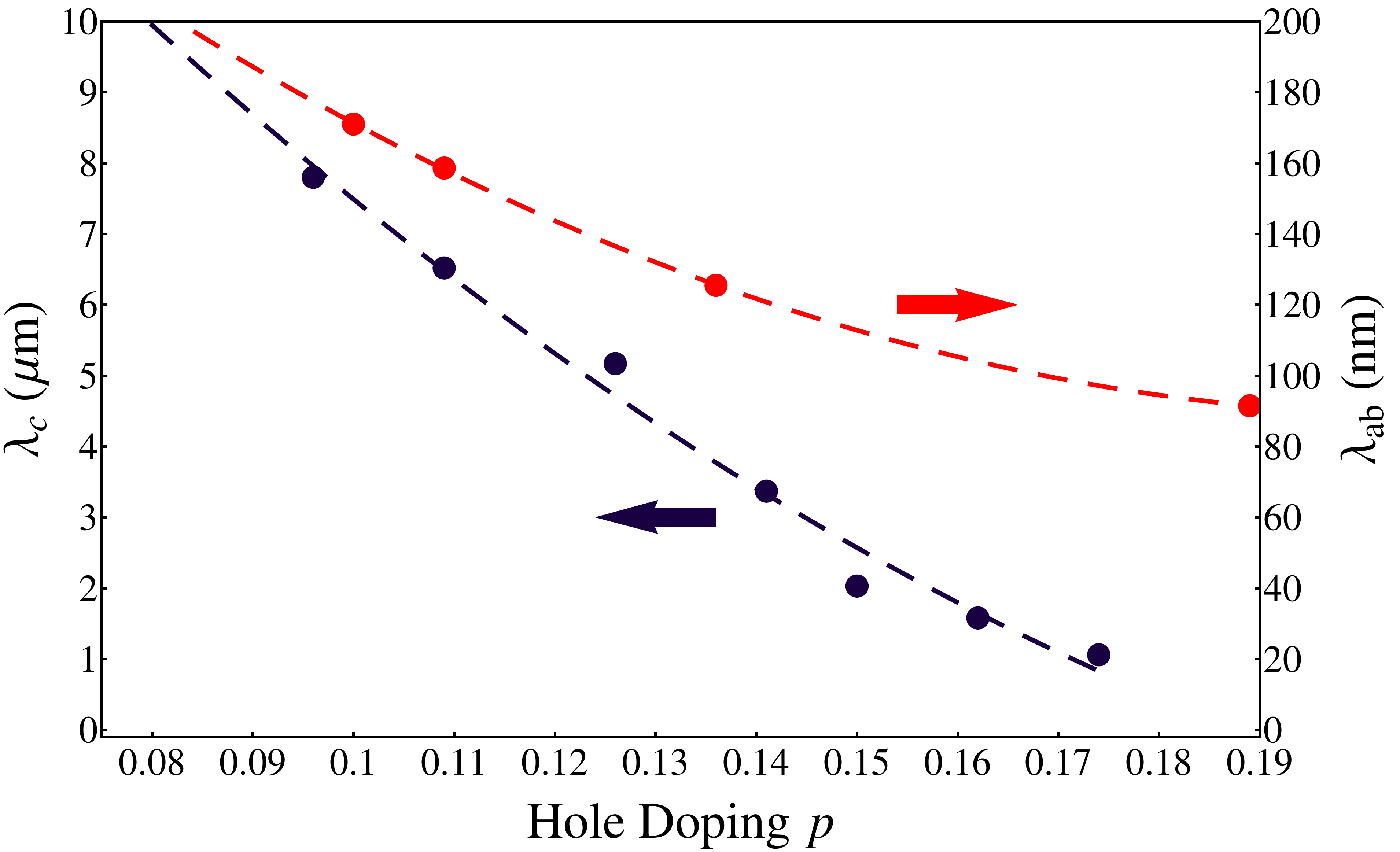}\\
\includegraphics[width=\columnwidth,natwidth = 38.81cm, natheight = 23.06cm]{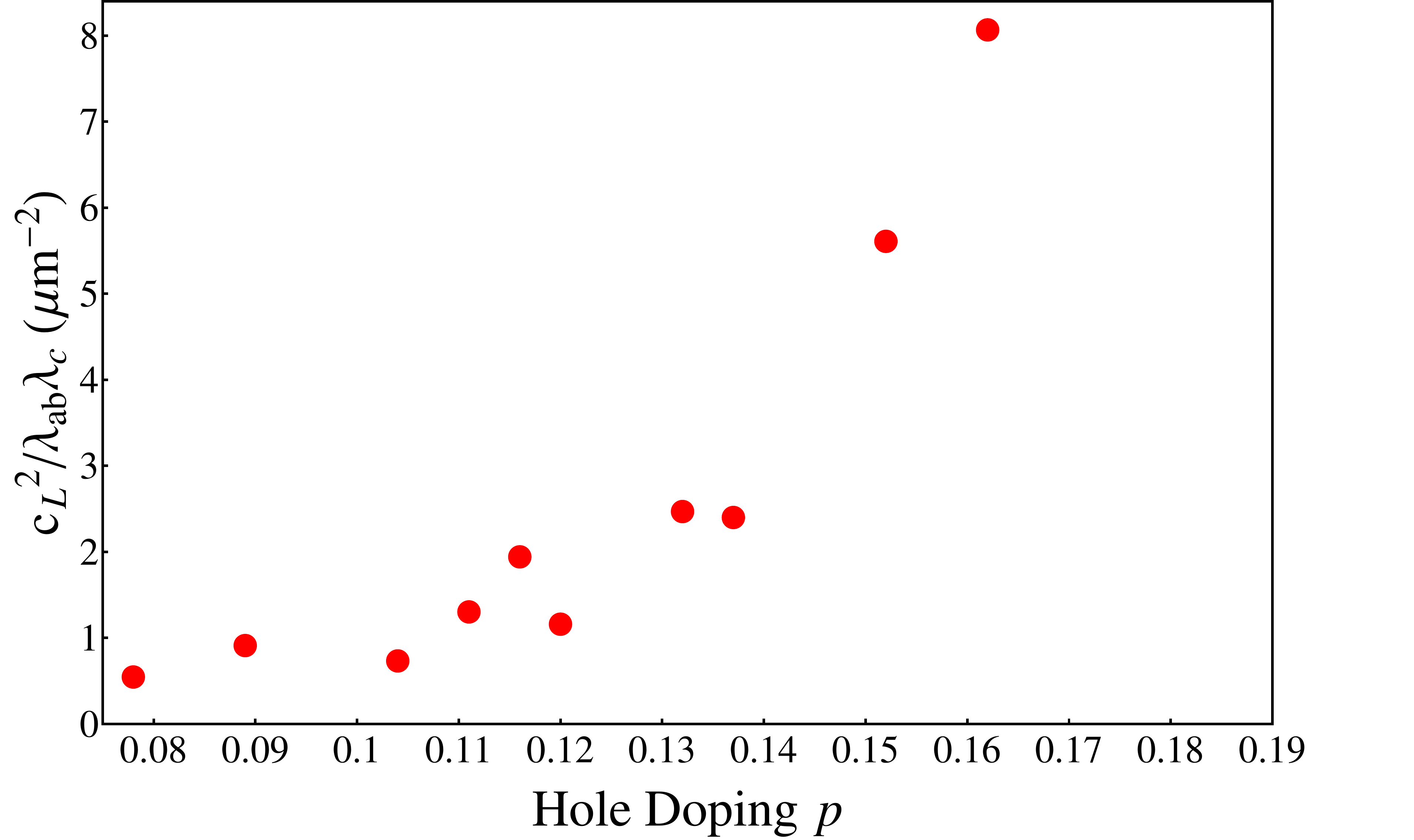}
\caption{\textit{Upper panel:} The in and out-of-plane penetration depths of \YBCOy, as measured by muon-spin rotation ($\lambda_{ab}$ at 0.11 holes), electron-spin resonance (the other $\lambda_{ab}$ points), and infrared reflectance ($\lambda_{c}$). \cite{Sonier:1997, Pereg-Barnea:2004, Homes:1995} Note the difference in scale for the left and right hand axes: the anisotropy is actually decreasing with increased hole doping.  The dashed lines are parabolic fits to the data, which are used to obtain interpolated values for doping levels not measured, and should be viewed as purely phenomenological. \textit{Lower panel:} The ratio of the Lindemann number squared to the product of the in and out-of-plane penetration depths.  This quantity, appearing on the right-hand side of \autoref{eq:bm1}, controls the curvature of \Bm vs. $T$.}%
\label{fig:penetration}%
\end{figure}

\autoref{eq:bm1} can be expanded about \Tc and solved for \Bm as shown in \citet{Blatter:1994}, but if the full temperature range from 1.5 K to \Tc is to be used then it is more accurate to fit to the full implicit expression for \Bm.  The use of \autoref{eq:bm1} requires both the in and out-of-plane zero-temperature penetration depths, as well as the \Tc: these values are also listed along with the hole doping (estimated using \citet{Liang:2006}) in \autoref{tab:correlation lengths}. The in-plane penetration depth values, $\lambda_{ab}$, come from electron-spin resonance (ESR) measurements \cite{Pereg-Barnea:2004} and from muon-spin rotation experiments, \cite{Sonier:1997} both performed on comparable \YBCOy crystals grown at UBC. In the case of the ESR values, the geometric mean of $\lambda_{a}$ and $\lambda_b$ was taken. Out-of-plane penetration depth values, $\lambda_c$, come from infrared reflectance measurements, \cite{Homes:1995} also performed on UBC crystals. Interpolated values for the penetration depth were used when the exact doping values were not available. The penetration depth values and the interpolation are shown in the upper panel of \autoref{fig:penetration}.

With $\lambda_c$, $\lambda_{ab}$, and \Tc experimentally determined, the data at each doping can be fit using only two parameters: $c_L$ and $\HcTwo(0)$. The fits in the lower panel of \autoref{fig:6p592plots} and in \autoref{fig:bvts} clearly show that three-dimensional vortex melting describes the in-field resistive transition in \YBCOy from $y = 6.45$ to $6.92$. The penetration depth anisotropy, $\gamma = \frac{\lambda_c}{\lambda_{ab}}$, changes from $\sim \!50$~at 6.45 to $\sim \!16$~at 6.92: this results in decreased curvature of the melting line as oxygen content (and hole doping) increases. This is the same behaviour seen in several different cuprates of varying anisotropy, reported in \citet{Ando:1999}. The $c_L$ and $\HcTwo(0)$ values extracted this way are given in \autoref{tab:correlation lengths} for all of the dopings measured.The fact that the Lindemann number remains relatively constant as a function of doping means that the shape of the melting curve is determined primarily by the penetration depths, which are becoming less anisotropic as hole doping increases. The Lindemann number and the penetration depths appear only as the ratio $\frac{c_L^2}{\lambda_{ab} \lambda_c}$ in \autoref{eq:bm1}, and we plot this ratio in the lower panel of \autoref{fig:penetration}. The increase of $\frac{c_L^2}{\lambda_{ab} \lambda_c}$ with hole doping is what is controlling the changing curvature as a function of doping.  With this parameter setting the shape, $\HcTwo(0)$ corresponds to the $T=0$ intercept of the melting curve.

It should be emphasized that in \autoref{eq:bm1}
\begin{equation}
\lim_{T \rightarrow 0} \Bm(T) = \muZero \HcTwo(0),
\label{eq:limit}
\end{equation}
and so the values of $\HcTwo(0)$ derived from fits to \autoref{eq:bm1} are determined mostly by the zero-temperature intercept of the data for  \Bm vs. $T$, and are essentially independent of the penetration depth values chosen. The penetration depths and the Lindemann number always enter \autoref{eq:bm1} as the ratio $\frac{c_L^2}{\lambda_{ab} \lambda_c}$, and so errors in the penetration depth values (which arise because we use interpolated values from the upper panel of \autoref{fig:penetration}) are absorbed into the fit value of $c_L$.

 In using \autoref{eq:bm1}, we have ignored the possibility that the onset of finite resistivity is due to the lattice depinning, and not actually melting. This assumption is probably justified, as the depinning transition is distinct from the melting transition in \YBCOy only for temperatures very close to \Tc and in samples with extremely low disorder. \cite{Liang:1996}  The fits shown in \autoref{fig:6p592plots} and \autoref{fig:bvts} show data up to near \Tc when available, but only data at temperatures less than $0.8\times\Tc$ were used in the fits (which also avoids any possible effects of XY-critical phenomena near \Tc).\cite{Fisher:1991} Additionally, all of the samples in this study (except for possibly the \YBCO{92} sample) have more disorder than the \YBCO{95} sample used in \citet{Liang:1996}. This is because the ortho-II, III, V, and VIII states are not perfectly ordered,\cite{Zimmermann:2003} and have more disorder than ortho-I ordered \YBCO{95}, which is close to stoichiometry. This disorder pushes the depinning transition closer to \Tc.


\begin{table}
\begin{center}
\begin{ruledtabular}
\caption{$\HcTwo(0)$ and $c_L$ as obtained by fitting the vortex lattice melting curves to \autoref{eq:bm1}. $\xi_0$ is calculated from $\HcTwo(0)$ using \autoref{eq:HcTwo}. The uncertainties come from the width of the resistive transition and the proximity of the lowest data point to $T = 0$~K. The hole doping is obtained from the \Tc, using Figure 3 of \citet{Liang:2006}. }
\begin{tabular}{p{11mm}p{9.5mm}lp{6mm}p{6mm}ccc}
  Oxygen  & Hole & \Tc& $\lambda_{ab}$ & $\lambda_{c}$  & $\xi_0$ &$\muZero\HcTwo(0)$& $c_L$ \\
	Content & Doping &   (K)  &   (nm)  &  ($\mu$m)  & (\AA)   & (tesla)& \\ 
	$y$ & $p$ & & & & & \\ \noalign{ \smallskip} \hline \noalign{ \smallskip}
	6.45 & 0.078 & 44.5 & 208 & 10.2 & $25.4\pm 0.5$& $50.8 \pm 2.0$  & 0.37  \\
  6.47 & 0.089 & 51 & 189 & 8.8 & $26.9 \pm 0.5$ & $45.2 \pm 1.6$  & 0.41 \\
  6.56 & 0.104 & 59 & 165 & 7.0 & $29.9 \pm 0.4$ & $36.9 \pm 1.0 $& 0.31\\
	6.59 & 0.111 & 61.5 & 155 & 6.3 & $34.3 \pm 0.3 $ & $28.0 \pm 0.3$ & 0.37\\
	6.67 I & 0.116 & 64.7 & 147 & 5.6 & $36.6 \pm 0.5$  &$24.5\pm 0.7$&  0.41 \\
	6.67 II & 0.120 & 66 & 144 & 5.3 & $36.1 \pm 1.5$ & $25.2 \pm 2.0$ &0.31 \\
	6.75 & 0.132 & 75.3 & 130 & 4.1 & $27.9 \pm 0.5$& $42.1 \pm 1.5 $& 0.37 \\
	6.80 & 0.137 & 80.5 & 125 & 3.7 & $27.0 \pm 0.9$ & $45.0 \pm 3.0$ & 0.34 \\
	6.86 & 0.152 & 91.1 & 111 & 2.4 & $21.5 \pm 0.6$ & $70.9 \pm 2.0$ & 0.39 \\
	6.92 & 0.162 & 93.5 & 104 & 1.7 & $16.4 \pm 0.8$ & $121.9 \pm 10.3$  & 0.38 \\
	\end{tabular}
\label{tab:correlation lengths}
\end{ruledtabular}
\end{center}
\end{table}

\begin{figure}%
\includegraphics[width=\columnwidth,natwidth = 38.81cm, natheight = 23.57cm]{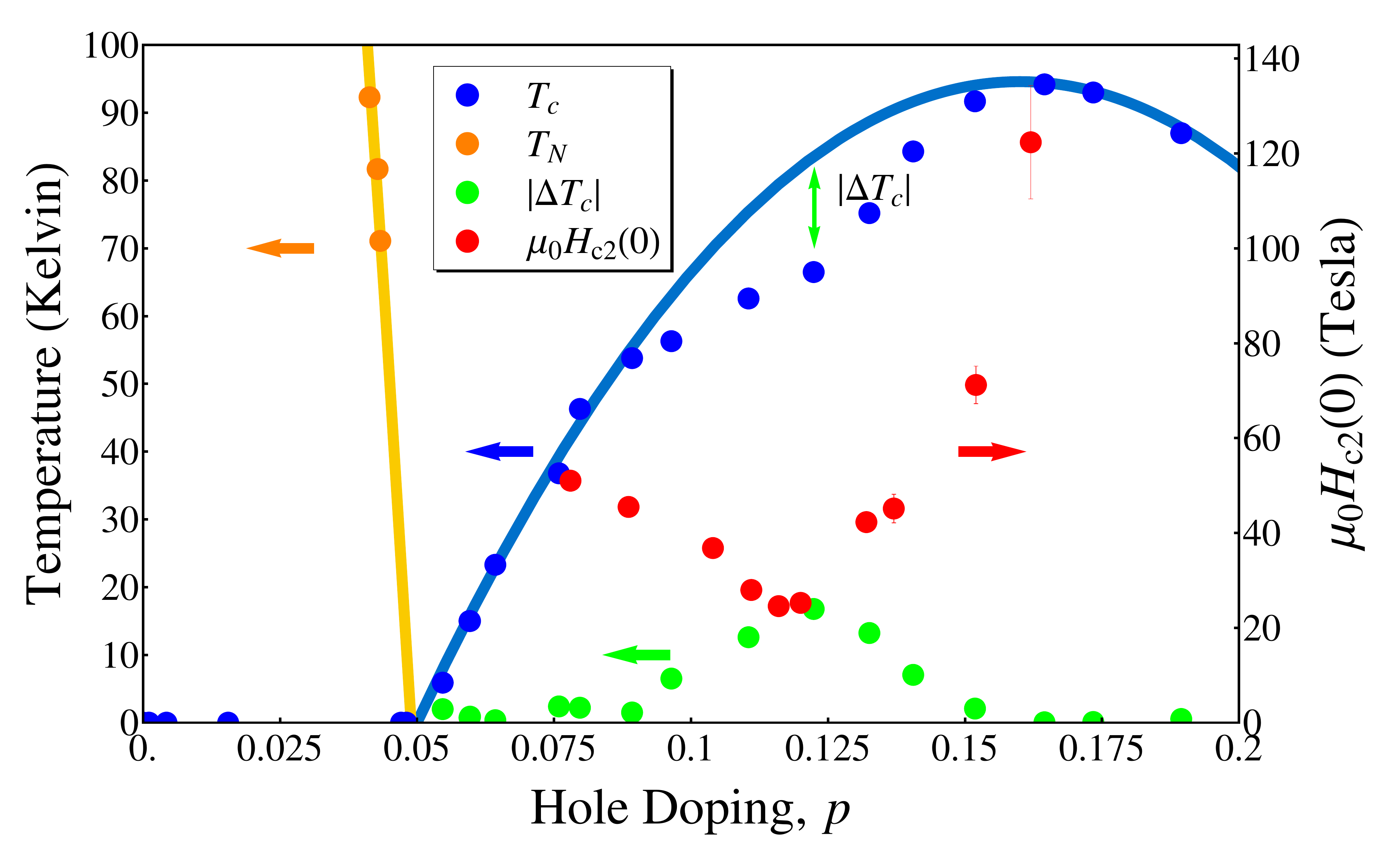}%
\caption{The superconducting phase diagram of \YBCOy. $\HcTwo(0)$ is suppressed in the same region that \Tc deviates from the parabolic form---a clear sign of the weakening of superconductivity around 0.12 hole doping. The \Tc values are taken from \citet{Liang:2006}.  }%
\label{fig:phasd}%
\end{figure}

The extracted values for $\HcTwo(0)$ are plotted with the phase diagram of \YBCOy in \autoref{fig:phasd}, and show an anomaly around 0.12 hole doping. The solid blue line in \autoref{fig:phasd} is the function
\begin{equation}
  1-\Tc / \Tc^{max} = 82.6 (p-0.16)^2,
\label{eq:tc}
\end{equation}
where $\Tc^{max}$ is the maximum \Tc of the material (equal to 94.3 K for \YBCOy).\cite{Liang:2006} This function has been found to describe \Tc as a function of $p$ in the cuprates, except for the suppression of \Tc around 1/8$^{th}$ hole doping. \cite{Liang:2006}  The green circles in \autoref{fig:phasd} are the absolute difference between the actual \Tc and \autoref{eq:tc}, and the suppression of \Tc is clearly correlated with a suppression of $\HcTwo(0)$. Suppression of the melting transition in this region was reported for a few different doping levels in \citet{LeBoeuf:2011}.

\section{Discussion}
The suppression of \Tc in the underdoped region of the phase diagram was mapped in detail by \citet{Liang:2006}.  In the same work, \citet{Liang:2006} correlated the \caxis lattice parameter with the hole doping of the copper-oxygen planes, showing a smooth evolution of hole doping with increased oxygen content. This demonstrates that the suppression of \Tc is not due to some peculiarity of the copper-oxygen chain doping mechanism in \YBCOy, but is in fact inherent to the electronic properties of the material. It was supposed that the suppression of \Tc may be due to a competition of superconductivity with stripe formation, as has been demonstrated explicitly in the lanthanum cuprates. \cite{Tranquada:1995}

The phase diagram in \autoref{fig:phasd} shows a clear correlation between the suppression of \Tc near 0.12 hole doping and a suppression in the $T=0$ melting field and hence a suppression of $\HcTwo(0)$. This further strengthens the case that the anomaly in \Tc is related to a weakening of superconductivity. The corresponding maximum in coherence length---recall that $\xi_0 \propto \left[\HcTwo(0)\right]^{-1/2} $---has also been seen in \muSR \cite{Sonier:2007} and in the fluctuation-magnetoconductance\footnote{In both of these experiments, the peak in coherence length was reported near oxygen 6.70, which would appear to be above where the peak occurs in \autoref{fig:phasd}. However, the oxygen content values reported may be shifted due to the different standards used for determining zero chain oxygen content, so that \Tc is a better indicator of the hole doping. \cite{Liang:2006} The peak in Figure 5 of \citet{Ando:2002b} appears at $\Tc = 60.5$~K, with the next data point at $\Tc = 69.5$~K.  The peak in \autoref{fig:phasd} of this current work occurs near $\Tc = 64.5$~K: these peaks are therefore consistent with each other. Similarly for \citet{Sonier:2007}, using \Tc as a means of computing hole doping reveals that their peak in $\xi_0$ corresponds to the same doping as our minimum in $\HcTwo(0)$.}. \cite{Ando:2002b} 

Recent NMR \cite{Tao:2011} and x-ray diffraction \cite{Ghiringhelli:2012, Chang:2012, Achkar:2012} experiments have indicated the possibility of charge order in underdoped \YBCOy. In all three x-ray diffraction experiments, the charge order was seen to drop in intensity below \Tc. Additionally, \citet{Chang:2012} found that the intensity of the charge-order peaks could be increased with an applied magnetic field below \Tc. These experiments give further evidence for a close competition between superconductivity and the charge ordered state. This is in agreement with the minimum in $\HcTwo(0)$ we observe near 0.12 hole doping.

\section{Conclusion}
The onset of finite resistivity in a magnetic field coincides with the vortex melting transition in \YBCOy. This melting transition can be substantially below mean-field \HcTwo at temperatures between 0 K and \Tc. \cite{Houghton:1989}  Using a Lindemann criterion for melting produces good agreement between theory and experiment, with a Lindemann number $c_L$ between 0.3 and 0.4. These values are consistent with theoretical predictions, which vary between 0.2 and 0.4 for highly anisotropic materials. \cite{Blatter:1994} Because this model agrees well with the data across such a wide range of dopings (and anisotropies) where $\HcTwo(0)$ is experimentally accessible, it is reasonable to assume that the extrapolations to zero temperature at higher doping levels gives a reasonable determination of $\HcTwo(0)$. 

Within the framework we used for flux-line-lattice melting, \cite{Houghton:1989, Fisher:1991, Blatter:1994} \Bm is required to approach $\muZero\HcTwo$ as $T \rightarrow 0$. The agreement between our data and this theory suggests that $\muZero\HcTwo(0) = \Bm(0)$, in contrast to previous suggestions. \cite{Wang:2002, Wang:2003, Wang:2006, Riggs:2011} This means that the quantum oscillations seen in underdoped \YBCOy would occur in a state free of vortices (superconducting fluctuations may still be present,\cite{Chang:2012b} of course, as detected in the Nernst signal,\cite{Chang:2010} for example.) This absence of vortices is consistent with the lack of a field-dependent scattering term needed to fully describe the quantum oscillations. \cite{Ramshaw:2011}

Below optimal doping, $\HcTwo(0)$ is rapidly suppressed with decreasing hole doping, reaching a minimum of 24.5 tesla at $p = 0.116$ holes. At lower hole doping $\HcTwo(0)$ recovers---even as \Tc continues to decrease---indicating the presence of a phase that competes with superconductivity, a phase which is strongest between 0.11 and 0.13 holes. 

\section{Acknowledgements}

The authors acknowledge support from the Canadian Institute for Advanced Research, the Natural Sciences and Engineering Research Council of Canada, the Canada Foundation for Innovation, and Euromagnet II. L.T. Acknowledges a Canada Research Chair and Fonds Qu\'eb\'ecois de la Recherche sur la Nature.
\newline
\bibliographystyle{apsrev4-1}
\bibliography{biblio}

\end{document}